\documentclass[useAMS,usenatbib]{mn2e}
\usepackage{subfigure}
\usepackage{natbib}
\usepackage{longtable,lscape}
\usepackage{threeparttable}
\usepackage{verbatim}
\usepackage{rotating}
\usepackage{color}
\usepackage{graphicx,epsfig}

\newcommand{\comments}[1]{}
\newcommand{\tco}{\mbox{$^{13}$CO(3-2)}}

\newcommand\mslin{M$_\odot$ pc$^{-1}$}
\newcommand\ms{M$_\odot$}
\newcommand\kms{ km s$^{-1}$}
\newcommand\kcm{ K cm$^{-3}$}

\newcommand\grcmt{g cm$^{-3}$}

\newcommand\cmdgr{cm$^2$g$^{-1}$}
\newcommand\meanp{$\langle P \rangle$}

\newcommand\ratiop{$P_s/\langle P \rangle$}
\newcommand\ratiom{$m/m_{vir}$}
\newcommand\mw{$\mathcal{M}/|\mathcal{W}|$}
\newcommand{\mum}{$\mu$m}

\bibpunct{(}{)}{;}{a}{}{,}

\title[Structure and radial equilibrium of filaments]{Structure and radial equilibrium of filamentary molecular clouds}

\author[Y. Contreras, J. Rathborne and G. Garay]{Y. Contreras$^{1,2}$\thanks{E-mail: Yanett.Contreras@csiro.au}, J. Rathborne$^{1}$ and G. Garay$^{2}$\\
  $^{1}$CSIRO Astronomy and Space Science, PO Box 76, Epping NSW 1710, Australia\\
  $^{2}$Departamento de Astronom{\'i}a, Universidad de Chile, Casilla
  36-D, Santiago, Chile}

\begin{document}

\date{Accepted xxx. Received xx; in original form xx}


\maketitle

\label{firstpage}

\begin{abstract}
Recent dust continuum surveys have shown that filamentary structures 
are ubiquitous along the Galactic plane.  While the study of their global properties 
has gained momentum recently, we are still far from fully understanding their 
origin and stability. Theories invoking magnetic field have been formulated to help explain the stability of filaments; however, observations are needed to test their predictions. In this paper, we investigate the structure and radial equilibrium of five filamentary molecular clouds with the aim of determining the role that magnetic field may play. To do this, we use continuum and molecular line observations to obtain their physical properties (e.g. mass, temperature and pressure). We find that the filaments have lower lineal masses compared to their lineal virial masses. Their virial parameters and shape of their dust continuum emission suggests that these filaments may be confined by a toroidal dominated magnetic field.
\end{abstract}


\section{Introduction}

Recent surveys of dust thermal continuum emission
  (e. g. ATLASGAL, \citep{Schuller}; BOLOCAM, \citep{bolocam}; HiGAL,
  \citep{hershel}) have shown that filamentary structures are
  ubiquitous along the Galactic plane. While filaments associated with well-known
  star-forming regions have been studied individually (i.e. those associated with Orion, \citep{orion-johnstone}; 
Taurus, \citep{taurus}; Lupus, \citep{lupus};  Ophiucus, \citep{ophiucus}), these new surveys have identified a large number of filaments along the Galactic plane which allow us to obtain a census of these type of molecular clouds.

Despite their ubiquity and importance as hosts to the earliest stages of high-mass stars and clusters (e.g. \citet{jackson-2010}), very little is known about the origin, stability and detailed physical properties of filaments and their embedded clumps.

A handful of recent works \citep[e.g.][]{fiege-pudritz-2000,Hernandez-2011} have focused on the identification of 
the mechanism that is responsible for the stability of filaments, 
and in particular whether magnetic fields play a role. 
Observationally, there is growing evidence that magnetic fields are associated with  
filamentary molecular clouds. \citet{alves-2008} undertook optical polarimetry
towards the diffuse gas in the Pipe nebula and found a large scale magnetic
field perpendicular to the main axis of the cloud. \citet{chapman-2011},
studying the polarization of background starlight, also found a magnetic field
perpendicular to the long axis of the filamentary region B213 in Taurus. These
results are interpreted as evidence that the gas and dust within the cloud have collapsed along the
field lines forming the filamentary clouds. In Orion,
\citet{poidevin-2010} measured the polarimetry of stars finding two components which they interpret to be a  helical magnetic field wrapping around the filament.

Theoretically, the radial equilibrium of filamentary molecular clouds can be described 
by treating filaments as isothermal cylinders and using either pure hydrostatic 
models \citep{ostriker-64} or those
  that included magnetic fields \citep{fiege-pudritz-2000,tilley-2004,fiege-2004}. Because they have different predicted values for the density distribution and virial parameters, detailed observations of the gas and dust within and around filaments can help distinguish between these models.

Thus, to determine the radial equilibrium of filamentary molecular clouds, and whether magnetic 
fields play a significant role, we utilize data from the APEX Telescope Large Area Survey (ATLASGAL) to identify a sample of filamentary molecular 
clouds. The continuum  
data from this survey combined with molecular line observations will allow us to 
determine the physical properties of these filamentary structures and to assess 
their radial equilibrium.

Sections \ref{sec:idenfil} and \ref{sec:obs} describe, respectively, the filaments selected and the observations performed. In Section \ref{sec:physicalproperties}, we calculate the physical properties of the filaments. In Section \ref{sec:stability}, we discuss the structure and radial equilibrium of the filaments and compare their properties with theories that explain the equilibrium of filamentary molecular clouds and in  Section \ref{summary}, we present a summary of the main points addressed in this paper.

\begin{figure*}
  \begin{center}
  	\subfigure[][]{
	    \includegraphics[trim = 8mm 115mm 5mm
    		115mm,clip,width=.99\textwidth]{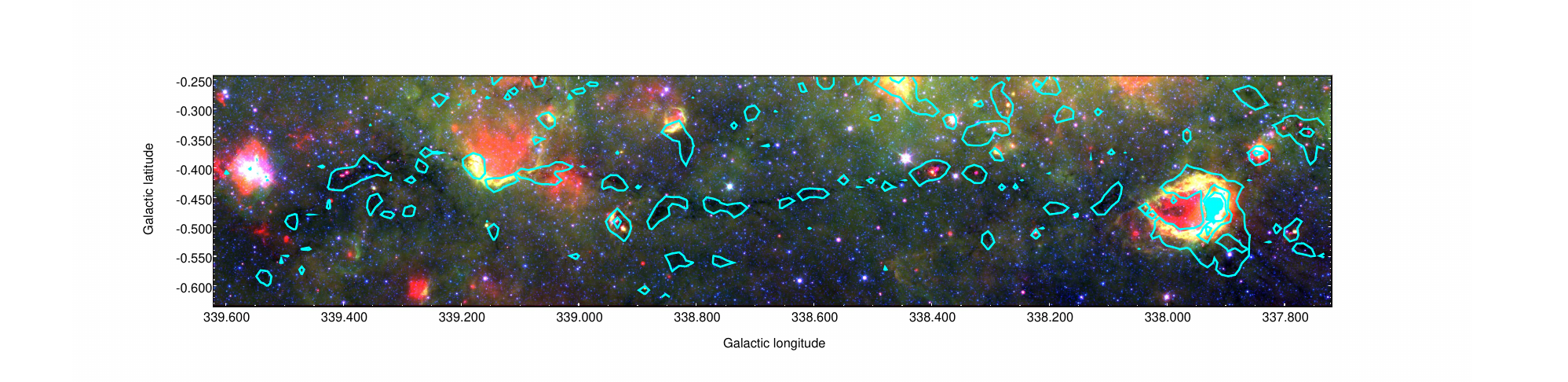}
    }\\
    \subfigure[][]{
     	\includegraphics[trim = 9mm 105mm 32mm
    		115mm,clip,width=.45\textwidth, height=4cm]{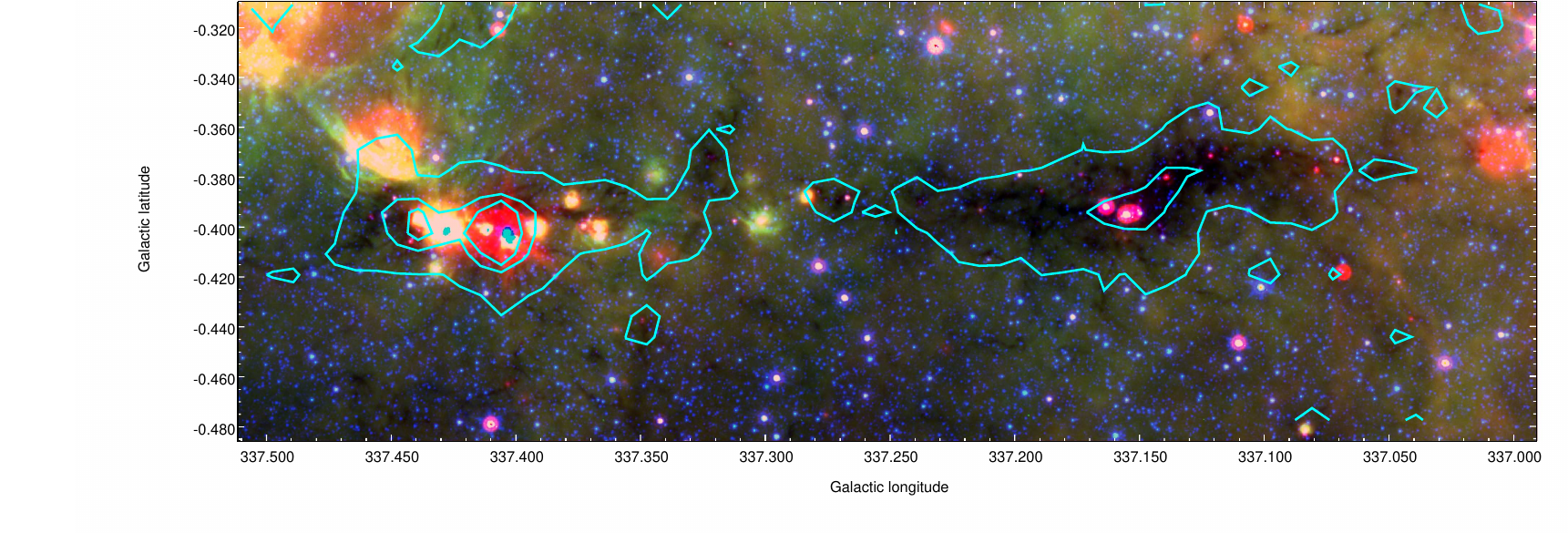}
    }
    \subfigure[][]{
 		\includegraphics[trim = 50mm 102mm 34mm
    		110mm,clip,width=.45\textwidth, height=4cm]{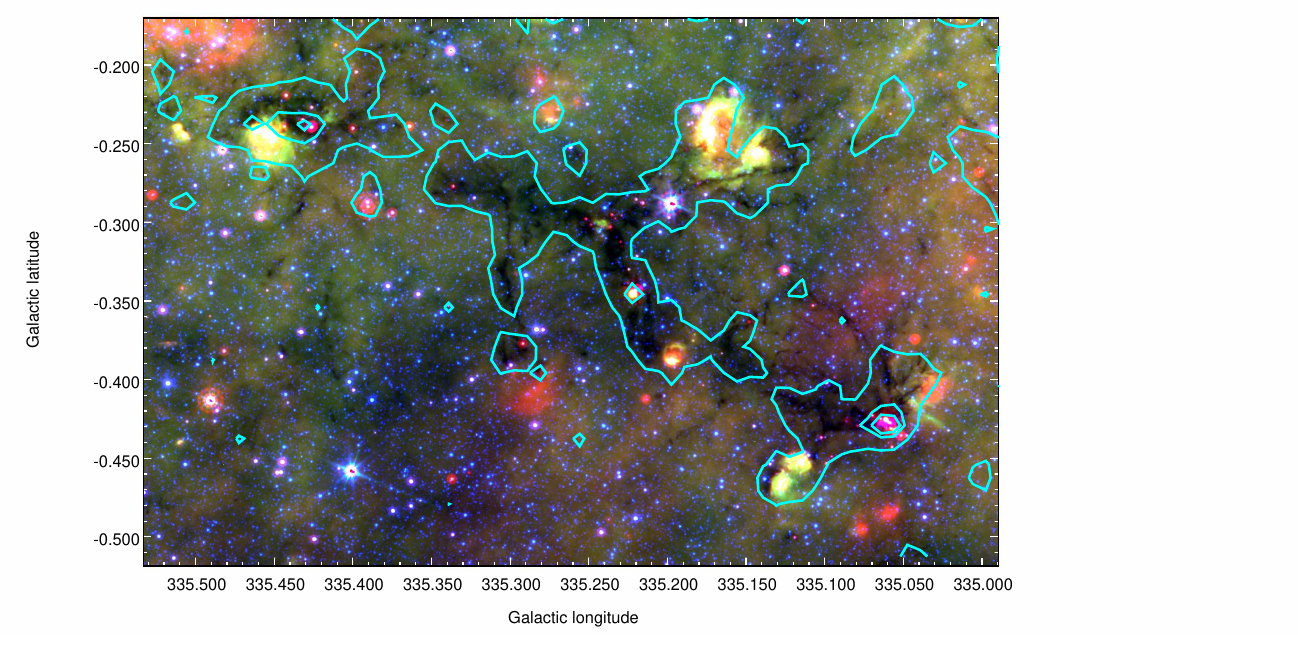}
    }
    \subfigure[][]{
 		\includegraphics[trim = 8mm 104mm 31mm
    		112mm,clip,width=8cm, height=4cm]{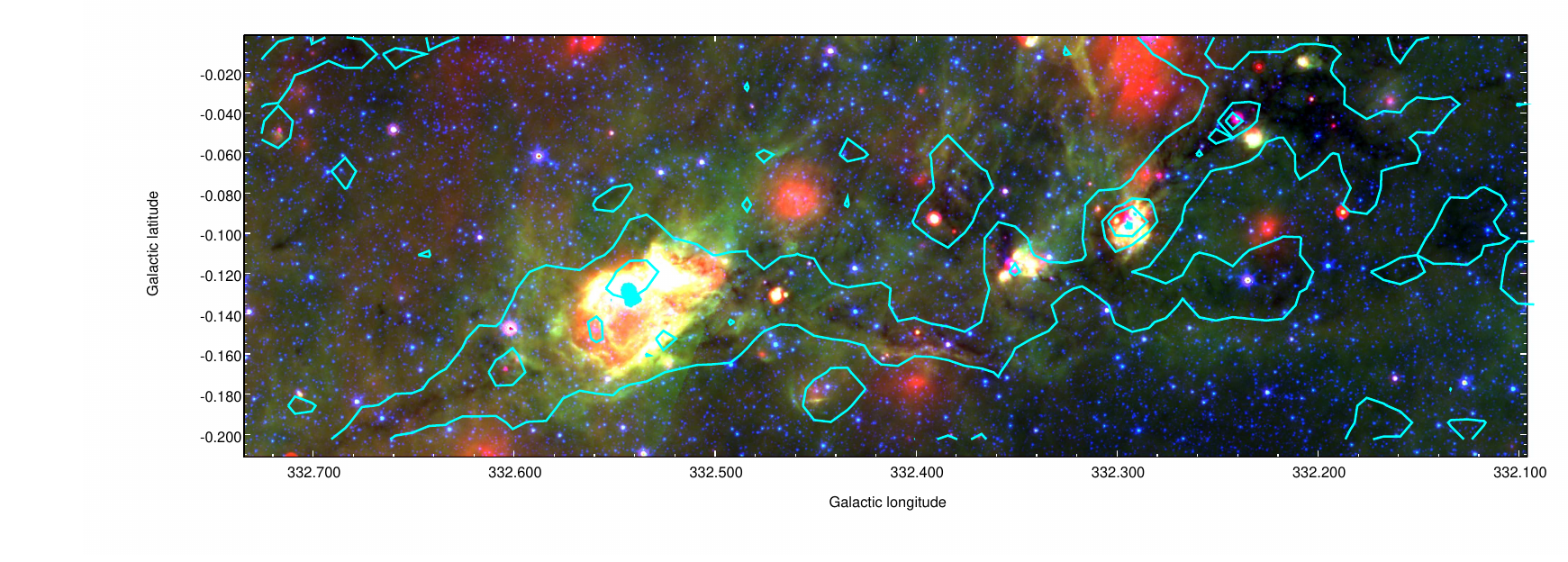}
    }
    \subfigure[][]{
 		\includegraphics[trim = 49mm 100mm 35mm
    		105mm,clip,width=8cm, height=4cm]{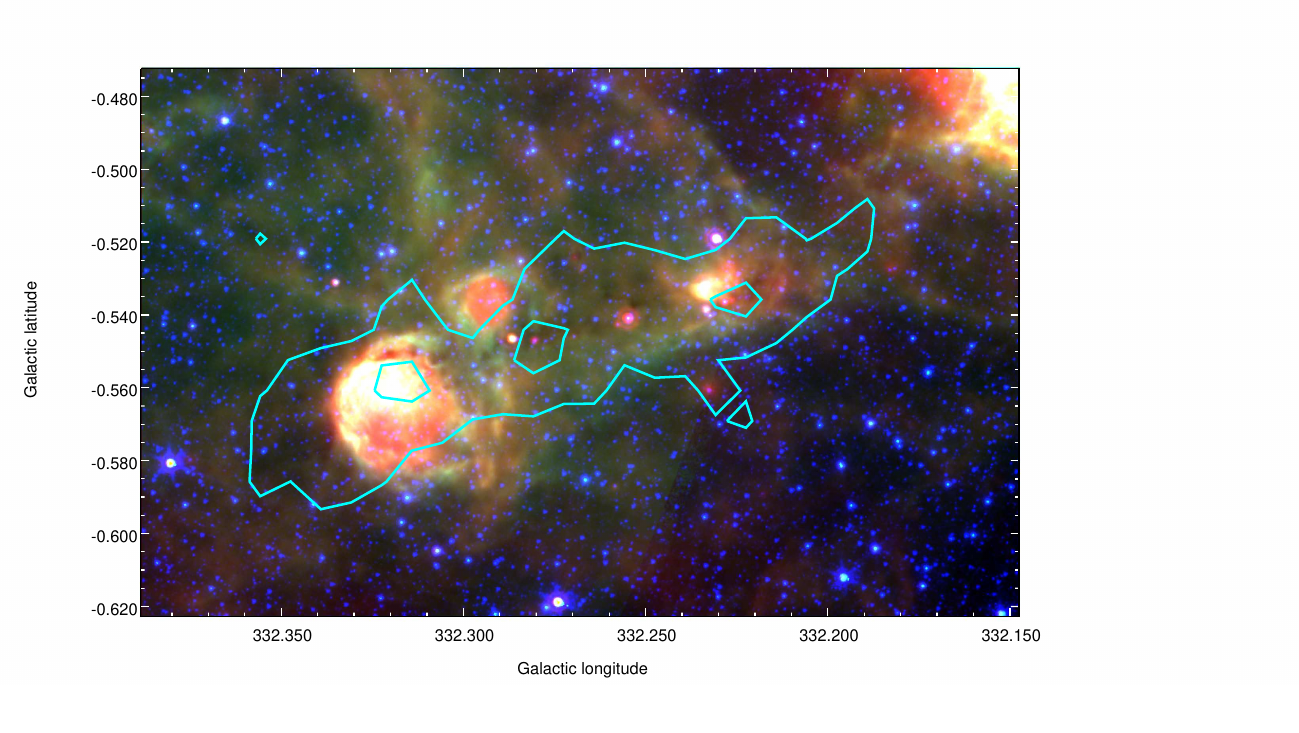}
	}
    \end{center}
  \caption{Three colour {\it Spitzer} images towards the five filaments selected overlaid with contours of the ATLASGAL 870~\mum~continuum emission (red: 24\mum~MIPSGAL; green: 8\mum~IRAC; blue: 4.5\mum~IRAC, contour levels are 3$\sigma$, 10$\sigma$ and 20$\sigma$). Panels (a) - (e) shows the images for each filament (A-E respectively). In all cases the filaments appear extended along the Galactic plane. Within them we see a range in star-formation activity, from infrared-dark to infrared-bright regions.}
  \label{fig:all-fil-color}
\end{figure*}

\section{Identifying filaments}\label{sec:idenfil}

To identify a sample of filaments, we selected a 20 square degree
region of the Galactic Plane centred at 335$^\circ$ Galactic
longitude. This region was selected because it contained many filamentary structures in the dust continuum and a mix of both dark and bright regions of infrared emission. Selecting filaments for further study that span a range in star formation activity, and presumably evolutionary stage, is important particularly when attempting to establish a connection between filamentary structures and the formation of high-mass stars and clusters.

We selected all filaments from the ATLASGAL 870~\mum~images. To enhance the extended emission associated with each filament, we first applied a Fourier filter to the images to remove the
small scale peaks. From this smoothed image (only including emission 
$>3\sigma$), we selected five structures that appeared to be connected. 
With this method we recovered the Nessie Nebula \citep{jackson-2010}, a unique filamentary 
structure recently identified as an infrared dark cloud (IRDC) from \textit{Spitzer} images. The 
potential filaments were named after the
brightest clump embedded within them, using the denomination from the
ATLASGAL compact source catalogue \citep{contreras-2013}. For simplicity we also assign them a short denomination (A,B,C, D and E).

We used {\it Spitzer} GLIMPSE \citep{benjamin-2003} and MIPSGAL \citep{carey-2005} images to characterize their associated infrared emission, which is an indication of their current star formation activity. Fig. \ref{fig:all-fil-color} shows the \textit{Spitzer} images overlaid with the dust continuum emission towards the selected filaments. This figure clearly shows the different infrared signatures associated with each filament. Parts of filaments `A: Nessie', `B: AGAL337.406-0.402' and `C: AGAL335.406-0.402' are seen as IRDCs, although some bright IR emissions are associated with several of their dense clumps.  These represent examples of filaments in a relatively early stage of formation.  Filaments `D: AGAL332.294-0.094' and `E: AGAL332.094-0.421'
show considerably brighter infrared emission and, thus, likely represent a more evolved state.

\section{Observations}\label{sec:obs}

\begin{table*}
\caption{Summary of the observation towards the filaments.}
\label{summ-obs}
\begin{center}
\begin{tabular}{c|ccccrcrrcc}
\hline\hline

&\multicolumn{2}{c}{Center of individual maps}&\multicolumn{3}{c}{Continuum (350~\mum)}&\multicolumn{3}{c}{$^{13}$CO(3-2)}\\
Filament&RA&Dec.&Size& rms &Date&Size&rms& Date\\
      &(J2000)&(J2000)& (arcmin) & \multicolumn{1}{c}{(Jy/beam)}&&(arcmin)&\multicolumn{1}{c}{(K)}&&\\
\hline\hline
A& 16:43:18.07&-46:37:31.50& 4.6x4.6&3.7& Oct 2011&4.6x4.6 &0.33&Jun 2010\\    
 & 16:43:05.61&-46:39:50.50& 5.0x5.0&4.4& Oct 2011&5.0x5.0 &0.35&Jun 2010\\    
B& 16:37:36.23&-47:41:35.55& 5.5x5.8&2.5& Aug 2010&5.4x5.8 &0.20&Aug 2010\\ 
 & 16:38:35.32&-47:30:15.21& 5.4x5.8&1.5& Aug 2010&5.3x5.9 &0.49&Aug 2010\\ 
 & 16:38:16.64&-47:33:52.82& 5.6x5.8&2.7& Aug 2010&5.7x5.9 &0.56&Aug 2010\\ 
 & 16:37:57.02&-47:37:47.75& 5.4x5.7&0.7& Aug 2010&5.5x5.5 &0.63&Aug 2010\\ 
 & 16:38:54.67&-47:26:33.94& 7.7x7.5&1.7& Aug 2010&7.7x7.5 &0.61&Jun 2010\\ 
C& 16:30:11.36&-48:48:02.33& 9.3x7.9&1.6& Jul 2010&4.9x10.6&0.44&Jul 2010\\ 
 & 16:29:52.64&-48:53:20.96& 7.5x9.1&0.9& Jul 2010\\                        
 & 16:29:54.61&-48:57:25.34& 5.3x5.8&1.0& Jul 2010\\                        
 & 16:29:35.92&-48:57:51.62& 5.7x6.1&1.0& Jul 2010\\                        
 & 16:29:36.95&-49:01:39.07& 5.7x6.2&1.1& Jul 2010&10.4x11.5&0.40&Jul 2010\\ 
 & 16:29:06.65&-49:00:33.41& 7.3x8.5&1.4& Jul 2010\\                        
 & 16:29:41.52&-49:05:42.80& 5.7x6.4&0.9& Jul 2010\\                        
 & 16:29:35.91&-49:09:01.85& 4.0x5.2&1.1& Jul 2010&4.5x6.6 &0.46&Jul 2010\\ 
D& 16:17:28.99&-50:46:18.41& 4.0x4.0&0.8& Oct 2011&4.0x4.0 &0.82& Oct 2011\\ 
 & 16:17:01.46&-50:48:01.81& 4.5x4.5&0.9& Oct 2011&4.5x4.5 &1.10& Oct 2011\\ 
 & 16:16:42.84&-50:50:16.61& 4.0x4.0&1.0& Oct 2011&4.0x4.0 &0.89& Oct 2011\\ 
 & 16:15:45.12&-50:55:57.62& 4.0x4.0&1.3& Oct 2011&4.0x4.0 &0.75& Oct 2011\\ 
 & 16:15:34.65&-50:55:42.83& 4.0x4.0&1.0& Oct 2011&4.0x4.0 &0.73& Oct 2011\\ 
 & 16:15:16.84&-50:55:59.55& 4.0x4.0&1.0& Oct 2011&4.0x4.0 &0.73& Oct 2011\\ 
E& 16:16:26.20&-51:17:45.83& 8.9x9.2&0.7& Aug 2010&8.9x9.3 &0.46&Oct 2010\\ 
 & 16:18:04.70&-51:08:27.65& 4.0x4.0&0.4& Oct 2011&5.5x5.5 &0.90& Oct 2011\\       
 & 16:17:56.11&-51:15:07.07& 5.5x5.5&0.5& Oct 2011&5.4x5.7 &0.34&Oct 2010\\   
 & 16:17:23.77&-51:17:10.16& 5.5x5.5&0.7& Oct 2011&5.9x5.3 &0.32&Oct 2010\\   

\hline\hline
\end{tabular}
\end{center}
\end{table*}

For all the filamentary structures we use the continuum data
at 870~\mum~from ATLASGAL \citep{Schuller} to determine
their dust properties. These data were complemented with observations of the
350 \mum~continuum and \tco~molecular line emission. The combination of these data will allow us to establish their
physical coherence and properties.

\subsection{ATLASGAL: 870 \mum~ continuum survey}\label{sec:870}

The ATLASGAL \citep{Schuller} mapped dust thermal emission at 870 \mum~ at
high sensitivity with the aim of obtaining a complete census of
regions of high-mass star formation within the Galaxy.

The survey was carried out from 2007 to 2010 using the Large APEX BOlometer CAmera
 \citep[LABOCA;][]{siringo-2009} mounted at the 12~m
sub-millimetre Atacama Pathfinder EXperiment (APEX) antenna
\citep{guesten}, located in Llano de Chajnantor, Chile. The survey
covered from $\pm 60^\circ$ in Galactic longitude, and $\pm 1^\circ.5$
in Galactic latitude. Also included in the surveyed region was the Carina arm, from  -80\degr~to -60\degr~in Galactic longitude and
-2\degr~to +1\degr in Galactic latitude.

LABOCA is a 295 element bolometer array centred at 345~GHz with a
bandwidth of 60~GHz, at this frequency the beam size of APEX is 19.2 arcsec. The errors in flux are
estimated to be lower than 15\% \citep{Schuller} and pointing rms is
$\sim$~4 arcsec. The final maps are generated with a pixel size of 6 arcsec,
corresponding to a 1/3 of the telescope beam. The rms of the maps is
typically 50 mJy/beam \citep[see][for further details]{Schuller}. More than 6000 compact sources have been identified within the Galactic plane coverage of this survey, with half of these sources  showing no obvious evidence for current star formation \citep{contreras-2013}.

\subsection{350~\mum~continuum emission}\label{sec:350}

The 350~\mum~continuum observations were obtained using the Submillimeter APEX
Bolometer Camera \citep{saboca}, which consists of a bolometer array
of 39 detectors, mounted on the APEX telescope. The beam size of APEX at this wavelength is 7 arcsec and the field of view
is $\sim$90 arcsec.  The observations were made using the on-the-fly (OTF) mode. To
avoid imaging artefacts due to the scan mode, each map was observed twice at orthogonal
scanning angles.

The observations were carried out during 2010 June, 2010 August and
2011 October. The images towards each filament were obtained by
the combination of several small adjacent maps centred along the length of each filament. Details of the individual maps for each filamentary structure (centre position, size and time) are summarized
in Table \ref{summ-obs}.

Pointing and focus checks were made using planets (Mars and Neptune
where available). The flux calibration was made using planets and secondary
calibrators with known fluxes: B13134, IRAS 16293, G5.89. Each flux calibration source
was observed every hour, and skydips were also performed every hour to
estimate the atmospheric opacity.

The data were reduced using {\tt BOA}\footnote{http://www.apex-telescope.org/bolometer/laboca/boa/} software designed to read, handle, and
analyse bolometer array data. Each small orthogonal map observed was reduced and calibrated independently. Each
map was opacity corrected, the base
line was fitted, and any point sources extracted. The noise in
the map was reduced with a median correlated noise filter and any
spikes removed. After this step the point sources were added back into
the maps and the fluxes were calibrated. Overlapping orthogonal maps were then
combined.

Once both orthogonal maps were added together, a signal to noise map
and a mask with the emission over a threshold of 5 $\sigma$ (with
$\sigma$ being the rms of the map) were constructed. Once we
obtained this mask, each map was re-reduced following the steps
mentioned above, using the mask as input for the emission. 

All the small maps were added together in {\tt BOA} to produce the larger
map that covered each filament. Since the individual maps
overlapped, the large map had a higher sensitivity than each
individual map. Thus, we created a new mask for each filament and all
the individual maps were re-reduced using this larger, more sensitive
mask to recover the extended emission. This reduction process was needed to best extract the complex extended
filamentary structures.

\subsection{$^{13}$CO molecular line emission}\label{sec:13co}

$^{13}$CO(3-2) observations across the filaments were made using the
APEX-2, superconductor-insulator-superconductor heterodyne
receiver at APEX during 2010 June, 2010 August, 2010 October and
2011 October. The beam size of APEX is 18.2 arcsec at the frequency of 
the $^{13}$CO(3-2) line. The
backend used during the 2010 observations was the Fast Fourier
Transform Spectrometer (FFTS) and the eXtended FFTS was used during the observations in 2011 October. The velocity resolution of the spectra is $0.07$~\kms.

Small regions across the filament were mapped using the OTF mode. Since $^{13}$CO is ubiquitous along the Galactic plane, for the signal-free off position we used
an absolute position below the Galactic plane to avoid contamination
in the spectra. Table \ref{summ-obs} gives the centre coordinates, the
size of the mapped regions, the average noise and date of the
observations.

The system temperature was calibrated prior to each observation using the
chopper mode. Pointing and focus were made on IRAS15194 every hour, the
pointing accuracy was  2-3 arc sec. The average value of the water vapour
column density ranged from 0.6 mm to 1.4 mm and the system temperatures were
typically 300~K. T$_A$ were converted to T$_{MB}$ using a mean beam
efficiency, $\eta_{MB}=0.73$.

We used Continuum and Line Analysis Single-dish Software to
reduce each of the individual maps. For each spectrum a
baseline was subtracted, using a three degree
polynomial. The data were then gridded and adjacent maps were then combined to produce large maps that covered each filament.

\section{Physical coherence and properties of the filaments}\label{sec:physicalproperties}

\subsection{Physical coherence}

Using the \tco~molecular line observations we are able to assess whether the 
filamentary continuum structures selected from the ATLASGAL maps correspond to a 
single physical structure or are the result of a projection effect from different 
clouds along the line of sight. For this, we use the $v_{LSR}$ measured along each filament to determine the continuity of the emission observed. We found that for all filaments, except B, the emission observed in 
the continuum images correspond to single physically coherent structure. 

We note that our observations of `Nessie' (filament A) cover only a small fraction of its 
length. The average velocity of the mapped region is $-37$ \kms, consistent with the 
results  of \mbox{HNC(1-0)} observations made by \citet{jackson-2010} towards the 
total extent of Nessie. 

For filament B we found that the dust continuum emission is associated with three discrete molecular clouds along the line of sight. A background cloud with average velocity of $-48$~\kms, which will be excluded for the analysis of this filament; an elongated molecular structure, having an average velocity of $-41$~\kms~that matches the morphology of the continuum emission, and which will be considered as the main filament; and a foreground cloud with an average velocity of $-18$~\kms~located towards one side of the filament. This component will be taken into consideration when calculating the physical properties of this filament.

For filament C the molecular line emission has an average velocity that ranges from $-40$~to \mbox{$-45$\kms}, suggesting that it correspond s to a single coherent molecular cloud. For filament D we found that the molecular line emission has an average velocity of $-48$~\kms, which we also interpret to arise from a single molecular cloud. For filament E, the emission has an average velocity of \mbox{$-53$~\kms}.

\subsection{Kinematic distance}

We used the line velocity obtained from the \tco~to derive their kinematic distances to each filament, using the rotational curve model described in \citet{brand-blitz}.

Filaments A, B and C are clearly seen in extinction against the bright diffuse 
background, therefore we assume they are at the near kinematic distance. 
While filaments D and E show both dark and bright IR emission, we also assume that 
they are at the near kinematic distance. The kinematic distances for the filaments 
are summarized in Table \ref{table-summ}.

\subsection{Dust temperature}\label{sec:filA-temperature}

We calculate the dust temperature or \textit{`color temperature'} of the filaments using the
ratio between the continuum emission at 870~\mum~and 350~\mum~using the expression \citep{Schnee-2005}:

\begin{equation}
  \label{eq:temperature}
  \frac{S_{870 \mu m}}{S_{350 \mu m}}=
  \left(\frac{870 \mu m}{350 \mu m}\right)^{3+\beta}
  \frac{\exp(h 350 \mu m/kT_{dust})-1}{\exp(h 870 \mu m/kT_{dust})-1},
\end{equation}

\noindent where $S_{870 \mu m}$ and $S_{350 \mu m}$ are the fluxes at \mbox{870 \mum}~and \mbox{350 \mum}~respectively, $h$ is the Planck constant, $k$ is the 
Boltzmann constant and $T_{dust}$ is the temperature of the dust. For this 
calculation we smoothed the 350 \mum~images to match the beam of the
870~\mum~data (18 arcsec). Using these wavelengths, the `colour temperature' will have 
sensible values towards regions of low temperatures (e.g. $<$20 K). Thus, it will 
be adequate for filaments that appear as IRDCs with little evidence of star 
formation.

Because the colour temperature depends on the value of the spectral
index $\beta$, we calculated the colour temperature using four
different values for $\beta$: 1, 1.5, 2 and 2.5.

In all filaments we found that the temperature increases towards the clumps, 
reaching in some clumps up to 100 K. In regions without obvious star formation, 
the average colour temperature derived for the filaments ranged from 8 to 20 K, 
depending on the value of $\beta$ used. For the following analysis a spectral 
index of $\beta=1.5$ was chosen. The temperatures listed in Table 2 correspond to the ones obtained using this value of $\beta$.

\subsection{Lineal masses along the filaments}

For all filaments we computed the mass per unit length (or lineal mass) from 
the dust continuum emission and the lineal virial mass from the \tco~emission. 
We divided each filament along its length in sections, d$x$, with a 
width equal to the beam size (18 arcsec) and computed the mass in each of these segments.

For each section of the filament we computed the 870 \mum~total flux 
by integrating the continuum emission across the diameter of the filament. From this, we computed the dust lineal mass, assuming a dust opacity of \mbox{$k_{870}=0.012$ \cmdgr} and used the temperature profile obtained from 
the dust temperature analysis, taking into account, in this way, the temperature variations along the filament. The average value of the lineal mass obtained for all the filaments ranged from $142$ to $400$ \mslin~(see Table \ref{table-summ}).

For each segment of the filament, we also obtained the lineal virial mass. 
To calculate this we averaged the molecular line emission across the diameter of the 
filament and performed a Gaussian fit to the averaged spectrum,
obtaining the central velocity, antenna temperature and line width. Then, the 
lineal virial mass was computed as
\begin{equation}
   m_{vir}=\frac{2\langle\sigma^2\rangle}{G},
\end{equation}

\noindent where $\langle\sigma^2\rangle$ is the average velocity dispersion 
in the segment and $G$ is the gravitational constant. $\sigma$ is determined from the line width using $\sigma=\Delta v_{tot}/\sqrt{8 ln(2)}$, where $v_{tot}$ is the observed line width plus a correction to take into account the mass of the observed molecule \citep{fuller-1992}.

Because the \tco~observations indicate that for part of filament B  the observed 
emission is actually the superposition of two discrete molecular clouds, the
lineal dust mass derived for this filament is likely to arise from the combination of
both molecular clouds. From the molecular line information, we can determine the virial mass of both molecular clouds. The foreground molecular cloud, at $-18$~\kms, has a mean line width of
$\Delta v=$ 1.3~\kms, which gives a mean lineal virial mass of \mbox{258 \mslin} and a total virial mass of $\sim10^3$ \ms. This correspond to $\sim$25\% of the total mass derived (foreground + filament) if we assume the two components were one coherent cloud. Therefore, for further analysis, we take this
contribution into consideration deriving a `corrected' mean lineal
dust mass and ``corrected" total dust mass for this filament.

For the filaments, the average value of the virial lineal mass ranged from 447 to \mbox{942 \mslin} (see Table \ref{table-summ}). We find that the average value of the virial lineal mass is higher than that of the lineal mass derived from the dust continuum emission,
which at face value suggests that the filaments are not in equilibrium and may be expanding. Hence, for the filament to be in virial equilibrium  in the absence of high external pressure, a magnetic field may be required to confine the filament.

\subsection{External and internal pressure}

To calculate the internal and external pressure, we consider both the region inside and outside the boundaries of the filament that are covered by the \tco~observations. The pressures
were computed via \citep{fiege-pudritz-2000}:

\begin{equation}\label{eq-apres}
  \langle P \rangle = \langle \sigma^2 \rangle \langle \rho \rangle,
\end{equation}

\noindent where $\sigma$ is the velocity dispersion and $\rho$ is the density. The mean density was obtained from the dust continuum emission while the velocity dispersion was obtained from the \tco~molecular line
emission.

We found that the average value for the line width in the regions outside the filaments ranged from  2.1 to 5.3 \kms. The corresponding densities ranged from $0.6$ to $7\times10^{-22}$ \mbox{g cm$^{-3}$}. Using these values we  derive external pressures ranging from $P_s/k=3.5\times10^4$ to $5.3\times10^5$ \kcm. These are
comparable to the values obtained for the external pressure towards 
other filamentary molecular clouds \citep[e.g. ][]{Hernandez-2011}.

The internal pressure was computed using the \tco~line width and the
density obtained for each point d$x$ along the filament. The density was obtained
by dividing the lineal mass by the volume per unit length ($\pi R^2$) at each point along the filament which was estimated using the derived radii, $R$.

The average line widths ranged from 1.8 to 3.3 \mbox{\kms}, and the average density found ranged from \mbox{$0.9 - 6.2\times10^{-19}$ \grcmt}. Using these values we derived internal pressures ranging from $\langle P\rangle/k=1.4\times 10^6 $ to \mbox{$2.5\times10^7$ \kcm}.

\begin{table*}
  \caption{Summary of the derived properties of the filaments from the observations at continuum and molecular lines.}
  \begin{center}
    {\small
    \begin{tabular}{cccccccccccccccc}
      \hline\hline
      {Fil.}&{Dist.}&Length&Radius
      &$\Delta v$&Temp.&$\Delta v_{ext}$&$\langle\rho_{ext}\rangle$
      &\multicolumn{3}{c}{Lineal mass}&$P_s/k$
      & $\langle P\rangle/k$\\
      
      &&&&&&&&{m}
      &$m_{vir}$&$\frac{m}{m_{vir}}$&&\\
      
      &(Kpc)&(pc)&(pc) &(km s$^{-1}$)&\multicolumn{1}{c}{(K)}&(km s$^{-1}$)&(\grcmt)&(\ms pc$^{-1}$)&(\ms pc$^{-1}$)&&(km s$^{-1}$)
      &(km s$^{-1}$)
      \\
      \hline\hline
      
      A&3.0&80 & 0.44 & 1.8&15&2.6&$1\times10^{-21}$& 262 &447&1.71& $10^{5}$  &1.8$\times 10^{6}$\\
      B&3.2&24 & 0.45 & 2.4&11&3.4&$3.5\times10^{-21}$& 410 & 552 &0.74& 5.3$\times 10^{5}$ & $10^{7}$\\
      C&3.6&33 & 0.56 & 2.7&11&2.1&$5.9\times10^{-21}$& 366 & 615 &0.46 \comments{0.28&1.14}&3.5$\times 10^{4}$  & 1.4$\times 10^{6}$&\\
      D&3.3&26.6&0.52 & 2.1&12&3.2&$3.7\times10^{-22}$& 291 & 380 &0.76 \comments{0.26&0.9}&4.9$\times 10^{4}$& 5.1$\times 10^{6}$ &\\
      E&3.7&36.6 & 0.23&3.3&12&5.3&$7.3\times10^{-22}$& 489\comments{& 492} & 942 &0.52& 2.8$\times 10^{5}$ &2.5$\times 10^{7}$ \\
      \hline\hline
    \end{tabular}
    }
  \end{center}
  \label{table-summ}
\end{table*}

\section{Structure and radial equilibrium of the filaments: the role of magnetic fields}\label{sec:stability}

In this section we investigate the structure and radial equilibrium of the filaments, assuming that they can be approximated as self gravitating cylinders. \citet{ostriker-64} studied the stability of isothermal self gravitating cylinders, deriving
analytical expressions for the distribution of pressure, density and
gravitational energy. In particular, they showed that
the density distribution, $\rho$, of an isothermal cylinder in equilibrium in the absence of a magnetic field can be
expressed as:

\begin{equation}
\label{eq:prof-nm}
\rho=\rho_0\frac{1}{\left(1+\frac{r^2}{8r_0^2}\right)^2},
\end{equation}

\noindent where $\rho_0$ is the central density, $r$ is the radius of the
filament and $r_0$ is the core radius of the filament. 

Several observation have, however, shown that large scale magnetic fields are present towards giant molecular
clouds \citep[e.g.][]{alves-2008, chapman-2011}. Thus, a more general theory to explain the
stability of filamentary molecular clouds should
include the presence of large scale ordered magnetic fields 
\citep{fiege-pudritz-2000,tilley-2004,fiege-2004}. \citet{fiege-pudritz-2000} found that when including magnetic fields the density profile falls off as $r^{-1.8}$ to $r^{-2}$, much shallower than that predicted by Equation 4 ($r^{-4}$).  This shallower slope agrees well with observations of filamentary molecular clouds
[e.g. L977 \citep{alves-1998} and IC5146 \citep{lada-1998}]. A density profile
of $r^{-1.8}$ to $r^{-2}$ is in agreement with the presence of a toroidal
magnetic field wrapping around the filament, which
prevents the filament from expansion \citep{fiege-pudritz-2000}.

In the \citet{fiege-pudritz-2000} model, the tensor equation of virial equilibrium is used to include the presence of a general helical magnetic field and an external pressure in long filamentary clouds. The virial equilibrium condition is then given by, 

\begin{equation}
  \frac{P_S}{\langle P
    \rangle}=1-\frac{m}{m_{vir}}\left(1-\frac{\mathcal{M}}{|\mathcal{W}|}\right),
   \label{eq:virial-fiege}
\end{equation}

\noindent where $P_s$ is the surface pressure; \meanp~ is the average internal
pressure, $m$ is the lineal mass; $m_{vir}$ is the lineal virial mass, $\mathcal{M}$ and $\mathcal{|W|}$ are the total magnetic and gravitational energies respectively given by:

\begin{equation}
  \mathcal{M}=\frac{1}{4\pi}\int{B_z^2d\mathcal{V}-\left(\frac{B_{zS}^2+B_{\phi S}^2}{4\pi}\right)\mathcal{V}},
\end{equation}
\begin{equation}
  \mathcal{W}= -m^2G,
\end{equation}

\noindent where $B_z$ is the magnetic field component along the axis of the
filament and $B_{zS},B_{\phi S}$ are the component z and $\phi$ of the
magnetic field at the surface of the filament.

The value of $\mathcal{M}$ can be positive or negative, depending
on  the strengths of the poloidal and toroidal components of the helical magnetic field. If $\mathcal{M}>0$, then the magnetic field is poloidal dominated, if $\mathcal{M}<0$ then the magnetic field is toroidal dominated. If
$\mathcal{M}=0$, then the helical configuration of the magnetic field would have a neutral effect on the radial equilibrium of the filament, and the non-magnetic
form of the virial equation is recovered. 

The value of \mw~can be constrained
by measuring the values of \ratiom~and \ratiop~from
observations. \citet{fiege-pudritz-2000} computed the values of \mw~for seven
filamentary molecular clouds, using values of mass and pressure taken from the
literature. They found that most of these filaments have values of \mw$< 0$,
which corresponds to a dominant toroidal magnetic field.

\subsection{Radial density profile analysis}

\begin{figure*}
  \begin{center}
    \includegraphics[trim =10mm 60mm 17mm
      60mm,clip,width=7.5cm, height=6cm]{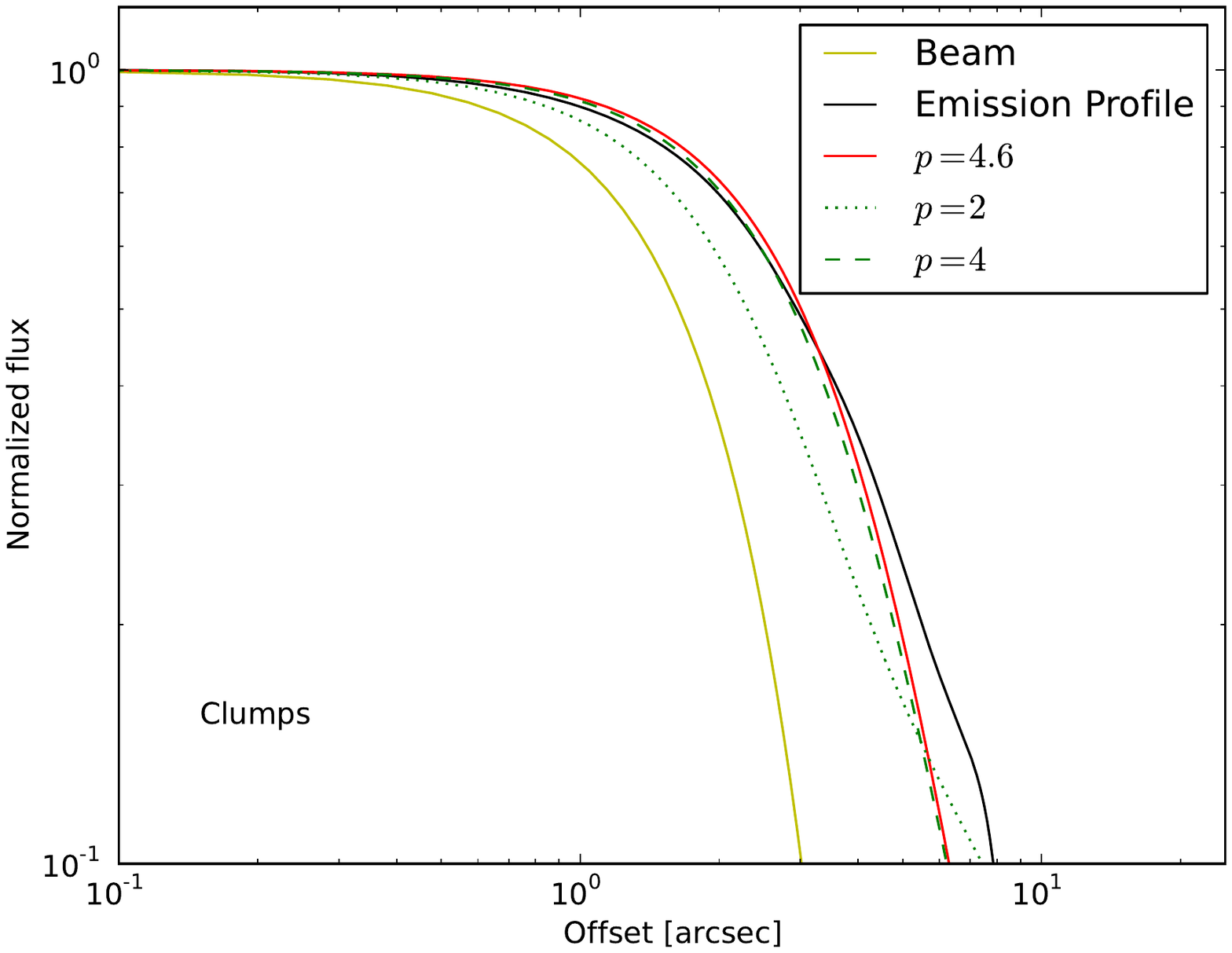}
    \includegraphics[trim =10mm 60mm 17mm
       60mm,clip,width=7.5cm, height=6cm]{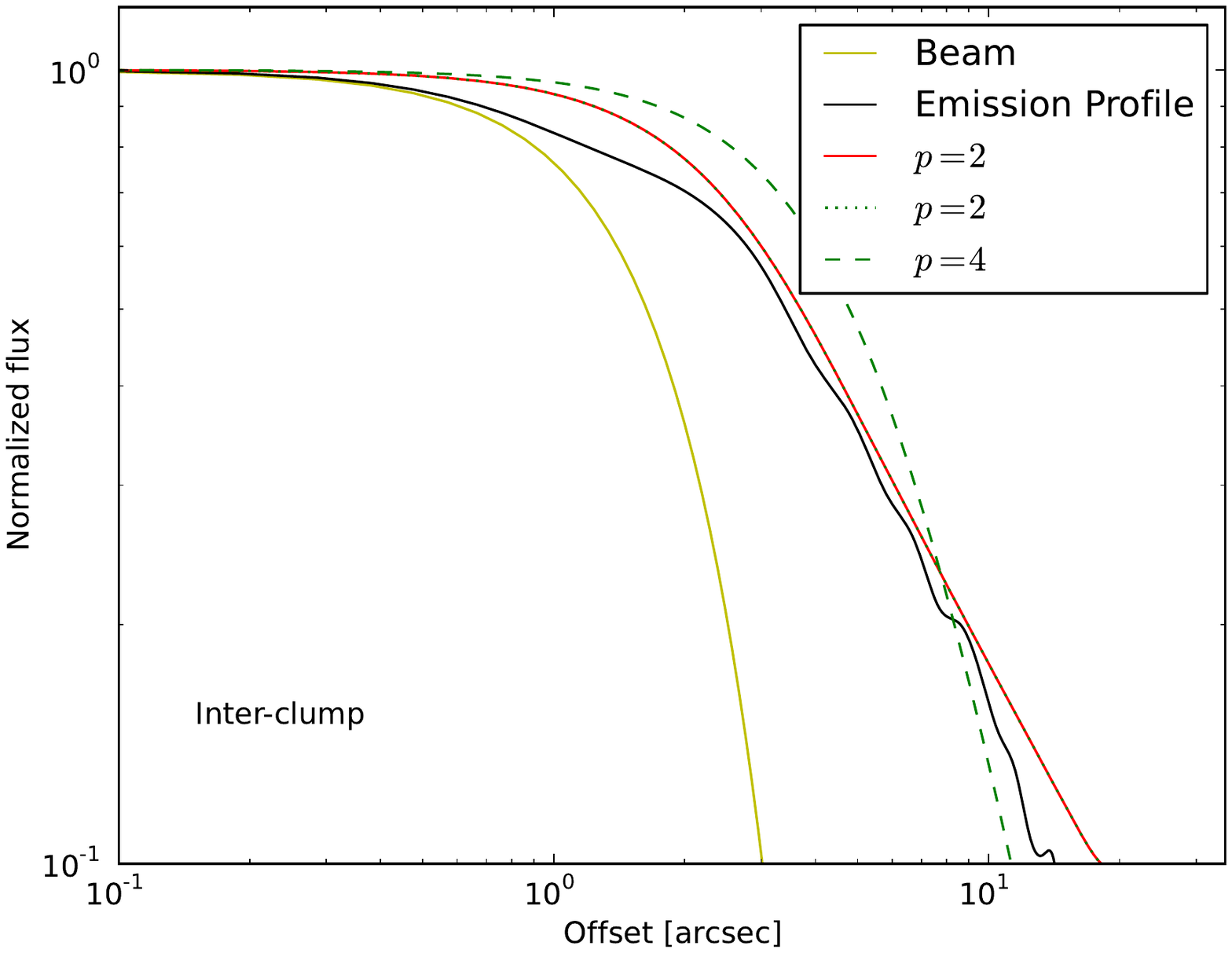}
  \end{center}
  \caption[Filament D: Density profile clumps]{Normalized intensity profile of the 870~\mum~emission from filament D (black line). The solid yellow line indicates the beam profile. The red solid line represents the best fit to the intensity profile. 
Left-hand panel: radial intensity profile of the emission from the clumps. Its shape can be well fitted by a profile with $p=4$ (hydrostatic solution; green dashed line). Right-hand panel: intensity profile of emission from the inter-clump material. 
Its shape can be well fitted by a profile with $p=2$ (magnetic solution; green dotted line).}
  \label{profile-clumps-interc}
\end{figure*}

In this section we investigate whether the presence of a magnetic field is necessary to explain the structure of the observed filaments, by analysing the profile of the radial intensity of the emission observed at 870 and 350~\mum. Theoretically, the intensity profile of a filament can be expressed as \citep{Arzoumanian}:

\begin{equation}
  \label{eq:lineal-prof-2}
  I(r)=A_p\frac{\rho_cR_{flat}}{[1+(r/R_{flat})^2]^{\frac{p-1}{2}}}\frac{k_\lambda B_\lambda(T_{dust})}{D^2},
\end{equation}

\noindent where $A_p$ is a finite number given by 
\mbox{$A_p=\int_\infty^\infty du/(1+u^2)^{p/2}$}; $\rho_c$ is the central density; 
$R_{flat}$ is the radius of the inner flat region; $k_\lambda$ is the dust opacity; 
$B_\lambda$ is the Planck function; $T_{dust}$ is the dust temperature; $D$ is 
the distance to the filament. The parameter $p$ represents the shape of the intensity profile, and gives an indirect measurement of the magnetic field
support over the filament. For a pure gravitationally bound filament without a
magnetic field, the radial intensity is best fitted by a profile with $p =
4$ \citep{ostriker-64} (see Equation \ref{eq:prof-nm}). 
In the case that the filament is confined by a magnetic field, then the radial 
intensity is best fitted by a profile with $p< 3$ (\citet{fiege-pudritz-2000}).  

To compare our observations to these theories, we first divided the
emission within the filament into two groups: a group containing the \textit{clumps} and
 a group containing the \textit{inter-clump} material. Since the shape of
the filament's radial intensity is only characterized by the value of
$p$ and $R_{flat}$, we averaged and normalized the radial intensity for each
group. Thus, we obtained for both the clumps and inter-clump material
a mean value for the radial intensity from the
870~\mum~and 350~\mum~emission. The radial intensity at each point along the filament was fitted using
Equation \ref{eq:lineal-prof-2}, for both the emission at 870~\mum~and 350~\mum, using $k_{870}=0.012$
, $k_{350}=0.073$ cm$^2$ g$^{-1}$, respectively, and the derived dust temperature.

Fig. \ref{profile-clumps-interc} shows an example
of the fit made to the emission towards the clumps and inter-clump
material in filament D (the rest of the filaments show similar intensity profiles). We found that for all the filaments there
is a clear difference between the profile obtained towards the clumps
and inter-clump material. Table \ref{summ-fits-prof} summarizes the values obtained for the
fit towards the clumps and inter clump material in each filament. The mean
values of the intensity profile for the clumps and inter-clump material
suggest that while the radial profile observed towards the clumps can be described by
hydrostatic equilibrium the inter-clump material cannot.

For the filaments, a non magnetized isothermal model does not describe well the shape observed in the dust intensity profile. Because an isothermal model with a toroidal dominated magnetic field reproduces well the observed profiles, we speculate that the filaments may be wrapped by a magnetic field which is confining them against expansion. While we cannot firmly conclude the existence of a magnetic field, the derived radial intensity profiles are suggestive of a toroidal dominated field.

\begin{table}
 \caption{Summary of the values of $p$ obtained for the fits towards the clumps and inter-clump material within each filament. This table shows that the clumps are best fitted by profiles with values of $p$ between 3.5 and 4, while the inter clump material is best fitted by profiles with $p\sim2$.}

  \begin{center}
    \begin{tabular}{l|cccc}
      \hline
      \hline
      Filament & \multicolumn{2}{c}{Clump} & \multicolumn{2}{c}{Inter clump}\\
      &870~\mum &350~\mum&870~\mum &350~\mum\\
      \hline
      \hline
      A&5.0&-&3.0&-\\
      B&3.5&4.8&2.0&2.0\\
      C&3.9&5.0&3.1&2.1\\
      D&4.5&4.0&2.0&2.1\\
      E&3.4&3.9&2.6&3.0\\
      \hline
      \hline
    \end{tabular}
  \end{center}
  \label{summ-fits-prof}
\end{table}

\subsection{Virial equilibrium analysis}

We also analysed the radial equilibrium of the filaments by studying the relationship
between their lineal mass, lineal virial mass, internal pressure, external
pressure, gravitational energy and magnetic field via Equation 
\ref{eq:virial-fiege}. Using the values obtained for each filament, we computed 
the values of \mw~(see Table 4). 

\begin{figure}
  \begin{center}
    \includegraphics[trim = 5mm 60mm 15mm
    60mm,clip,width=.45\textwidth]{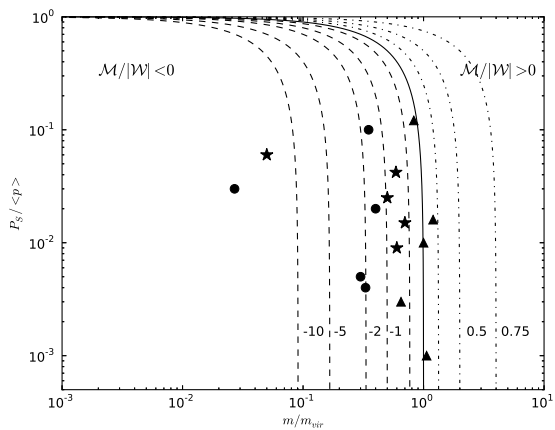}
  \end{center}
  \caption{\ratiop~versus \ratiom\ diagram. The lines indicate model results. Dashed: toroidal magnetic fields; solid: solution with neutral magnetic field, including the unmagnetized solution; dotted: poloidal magnetic field. The symbols indicate the derived values for our filaments. Stars: average value for each filament; dots: values for clumps; triangles: values for the inter-clump material.}
  \label{fig:a-fiege}
\end{figure}

Fig. \ref{fig:a-fiege} shows a \ratiop~versus \ratiom\ diagram 
for different values of the ratio between the total magnetic energy
per unit length, $\mathcal{M}$, and lineal gravitational energy,
$|\mathcal{W}|$. 
Also shown are the values obtained for the clumps (triangles), 
the inter-clump material (dots) and the average value of each filament (stars).
For the inter clump material, we found values of 
\mw~ranging from -1.3 to -34 suggesting the presence of a toroidal magnetic 
field. The average values of \mw~for the whole filaments 
range from -0.3 to -8.4 also suggesting the presence of a toroidal dominated 
magnetic field.
 
For the clumps, we found that the average value of \mw~ranges between 
$-0.5$ and 0.2, suggesting that in these structures the role of the magnetic fields  
are not very important. For this calculation, to compare directly the clump and inter-clump material, we use the same form of the virial equilibrium used to describe a filament. Because the clumps are probably spherical, this assumption may not be appropriate. Nevertheless, the fact that our virial equilibrium analysis suggest a toroidal magnetic field towards the inter-clump material but neutral in the clumps may be the result of the processes that lead to the formation of clumps within the filaments.

This virial equilibrium analysis suggests that if we treat the
filament as a whole coherent structure, the values obtained for their lineal mass, virial lineal mass, external and internal pressure suggest that their equilibrium can be explained by the presence of a toroidal dominated magnetic field that prevents the
filament from expansion. This result is also consistent with the analysis
of the radial density profile which suggest that the filaments need a
magnetic field to explain their radial equilibrium. 

\begin{table*}
  \label{summ-virial}
  \caption{Summary of the values of \ratiop, \ratiom and \mw for all the filaments.}
   \begin{center}
    \begin{tabular}{l|rrr|rrr|rrr}
      \hline
      \hline
      Filament &\multicolumn{3}{|c}{Clump} & \multicolumn{3}{|c}{Inter clump}&\multicolumn{2}{|c}{Average}\\ 
      &\ratiop &\ratiom &\mw &\ratiop &\ratiom &\mw&\ratiop &\ratiom &\mw\\
      \hline
      \hline
      A&0.121&0.8&-0.1&0.03&0.03&-34.0&0.06&0.1&-8.4\\
      B&0.020&1.2&0.2&0.10&0.4&-1.6&0.05&0.7&-0.3\\
      C&0.001&1.1&0.1&0.004&0.3&-2.0&0.025&0.6&-0.7\\
      D&0.010&1.0&0.1&0.02&0.4&-1.3&0.009&0.8&-0.3\\
      E&0.003&0.7&-0.5&0.005&0.3&-2.0&0.012&0.5&-0.9\\
      \hline
      \hline
    \end{tabular}
  \end{center}
\end{table*}

\section{Summary}\label{summary}

We used dust continuum data at 870 and 350~\mum~and molecular line data from five filaments in order to derive their properties and study their structure and radial equilibrium. Our main result and conclusions are summarized as follow.

A comparison of the lineal dust masses and their lineal virial masses reveal, in all five filaments, lower values for their lineal dust masses,
suggesting that an additional mechanism to gravity is needed to prevent these
filaments from against expansion. 

The radial intensity profile of
the dust emission at 870~\mum~and 350~\mum~are similar. We find that the radial intensity profile of the dust emission from regions containing clumps and inter-clump material exhibit different shapes. Towards the regions where the clumps are
located the radial intensity profiles are well fitted by a theoretical profile with
an index $p=4$. This profile describes the classical model of an isothermal, non-magnetic, cloud
in hydrostatic equilibrium \citep{ostriker-64}.

On the other hand, the radial intensity profile towards
the inter-clump material are  well fitted by a profile with an index $p=2$, which describes an isothermal filament, with magnetic field confinement. 

Considering the whole filament (i.e. inter-clump and clump region), the mean intensity profiles are well fitted with models with a radial density of $r^{-2}$ (i.e. $p=2$), suggesting that filaments are well represented by models where magnetic fields are needed for their radial equilibrium. This type of profile has also being observed in other filamentary molecular clouds
reported in the literature such as IC5146 \citep{lada-1999} or L977
\citep{alves-1998}.

When determining the stability and magnetic field support of the filaments via a
virial equilibrium analysis, we find that the average values of \ratiop~and \ratiom~for the filaments ranged from:

\begin{eqnarray}
0.5<&m/m_{vir}&<0.8\\
0.009<&P_s / \langle P\rangle&<0.06,
\end{eqnarray}

\noindent which implies a ratio between the lineal total magnetic energy
($\mathcal{M}$) and the lineal gravitational energy $\mathcal{W}$, of
\mw$<0$. This suggests that for the filaments to be in equilibrium a toroidal dominated magnetic field is required. This result is
similar to that found by \citet{fiege-pudritz-2000} towards the $\rho$
Oph, Taurus, and Orion regions. 

In summary, the analysis of the radial column density and of the
virial equilibrium suggests that filaments are best modelled
as isothermal filamentary
molecular clouds wrapped by a helical, toroidal dominated magnetic
field. This magnetic field would confine the filamentary molecular
cloud preventing its expansion and potentially influencing the formation of 
clumps within it. One suggestion is that the clumps might form along the
filaments at the location where the magnetic field crosses the
filaments; however, direct measurements of the magnetic field would be
useful to corroborate this hypothesis.

\section*{Acknowledgements}
We thank the referee Jason Fiege for his valuable comments that significantly improved this paper. Y.C. and G.G. gratefully acknowledge support from CONICYT through projects FONDAP No. 15010003 and BASAL PFB-06.

\bibliography{bibliografia}

\end{document}